\documentclass[%
 reprint,
 superscriptaddress,
 amsmath,amssymb,
 aps,
 prl,
]{revtex4-1}

\usepackage{graphicx}% Include figure files
\usepackage{dcolumn}% Align table columns on decimal point
\usepackage{bm}% bold math
\usepackage{gensymb}% Special symbols
\begin{document}

\preprint{APS/123-QED}

%---------------------------------------- Title and Authors ----------------------------------------
%===================================================================================================

\title{Thermally induced antiferromagnetic exchange in magnetic multilayers}

\author{D.~M.~Polishchuk}
\email{dpol@kth.se.}
\affiliation{Nanostructure Physics, Royal Institute of Technology, Stockholm, Sweden}%

\author{Yu.~O.~Tykhonenko-Polishchuk}
\affiliation{Nanostructure Physics, Royal Institute of Technology, Stockholm, Sweden}
\affiliation{Institute of Magnetism, NAS of Ukraine, Kyiv, Ukraine}

\author{E.~Holmgren}%
\affiliation{Nanostructure Physics, Royal Institute of Technology, Stockholm, Sweden}%

\author{A.~F.~Kravets}
\affiliation{Nanostructure Physics, Royal Institute of Technology, Stockholm, Sweden}
\affiliation{Institute of Magnetism, NAS of Ukraine, Kyiv, Ukraine}
 
\author{V.~Korenivski}%
\affiliation{Nanostructure Physics, Royal Institute of Technology, Stockholm, Sweden}%

\date{\today}% It is always \today, today,

\begin{abstract}

We demonstrate sharp thermally-induced switching between ferromagnetic and antiferromagnetic RKKY exchange in a spin-valve with the spacer incorporating a thin diluted ferromagnetic layer as the core. We illustrate the mechanism behind the effect as due to a change in the effective thickness of the spacer induced by the Curie transition into its paramagnetic state. 

%\begin{description}
%\end{description}
\end{abstract}

\maketitle

%\tableofcontents

%\begin{figure}[b]
%\includegraphics{fig_1}% Here is how to import EPS art
%\caption{\label{fig:epsart} A figure caption. The figure captions are
%automatically numbered.}
%\end{figure}

%---------------------------------------- I - Introduction -----------------------------------------
%===================================================================================================

The discovery of the indirect exchange coupling (IEC) of type Ruderman-Kittel-Kasuya-Yosida (RKKY) \cite{r1} and the giant magnetoresistance effect \cite{r2,r3} in magnetic multilayers have broadened a number of fields of physics and technology \cite{r4}. The discovered IEC oscillates in magnitude and sign versus the spacing of the individual ferromagnetic layers in a metallic stack \cite{r5}, yielding either parallel or antiparalel magnetic ground state of the multilayer, which is well explained theoretically as due to spin-dependent reflections and interference of conduction electrons within the nonmagnetic spacers \cite{r6,r7,r8,r9,r10}. This classical RKKY interaction is essentially independent of temperature \cite{r8,r11,r12,r13} and largely insensitive to any other external control parameter post-fabrication, which limits the use of the IEC in applications. The effect of alloying the spacer with nonmagnetic \cite{r16,r17,r18} and magnetic impurities \cite{r19,r20,r21} on RKKY was studied and explained in terms of an impurity-modified Fermi-surface topology and the corresponding significant changes in the RKKY-oscillation periods. The magnetic state of the spacer and its dependence on temperature was, however, was not investigated. Skubic et al. \cite{r21} reported on the competition between antiferromagnetic (AFM) RKKY exchange and direct ferromagnetic exchange interactions in Fe/V/Fe multilayers, where the spacer (V) was uniformly alloyed with Fe, but did not discuss the effect of temperature on the competing interactions in the system. 

Recent attempts to enhance the thermal effect on RKKY and use it to control the IEC in Tb/Y/Gd \cite{r14} and Co/Pt \cite{r15} multilayers focused on thermally affecting the properties of the respective softer ferromagnetic layers (Gd and thin Co) and thereby the spin-dependent reflection at the respective ferromagnetic interfaces (Gd/Y and Co/Pt). Both studies report relatively weak RKKY, without direct FM-to-AFM thermal switching of the magnetization, with relatively broad thermal transitions (of the order of 100~K, to near full strength RKKY). 

Here, we focus on thermally altering the effective spacer thickness and demonstrate a magnetic phase transition in Fe/Cr-based multilayers with gradient-doped spacers from strongly ferromagnetic RKKY at low-temperature to strongly antiferromagnetic RKKY at high temperature, both of the order of 100~mT in strength. By optimizing the material system and tailoring the mechanism involved, which is principally different from the previous studies, we achieve direct and fully reversible thermal switching of the RKKY interaction, from strongly ferromagnetic to strongly antiferromagnetic, with very narrow transition widths, of the order of 10~K, essentially in any desired temperature range, including room temperature. These results add efficient tuneability to IEC in magnetic nanostructures already used in spintronics, which should be highly technological in terms of suitable materials and operating field-temperature regimes. 

%##################### Figure 1 #####################
%===================================================
\begin{figure}[b]
\includegraphics[width=0.7\linewidth]{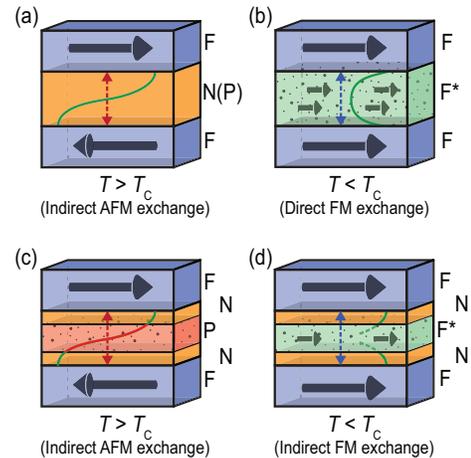}
\caption{Illustration of magnetization states in F/S/F trilayers with (a) uniform nonmagnetic (N) or paramagnetic (P) spacers, S = N(P); (b) uniform ferromagnetic spacer, S = F*; (c) composite N/P/N; and (d) composite N/F*/N spacers. Curie transition in F* within the composite spacer (transition from case (c) to case (d) on increasing temperature) changes the sign of RKKY as a result of a change in the effective thickness of the composite spacer.}
\label{fig_1}
\end{figure}
%===================================================

%##################### Figure 2 #####################
%===================================================
\begin{figure}%[c]
\includegraphics[width=0.8\linewidth]{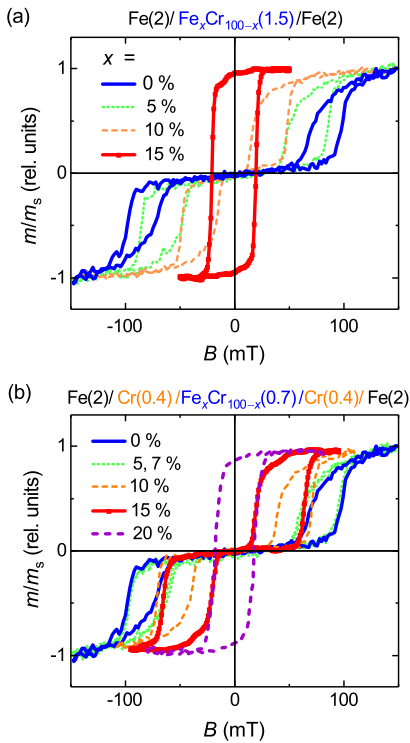}
\caption{Room-temperature VSM M-B loops for Fe/S/Fe trilayers with (a) uniform spacer S~= Fe$_x$Cr$_{100-x}$(1.5) and (b) composite spacer S~= Cr(0.4)/Fe$_x$Cr$_{100-x}$(0.7)/Cr(0.4), for different Fe-concentrations, $x$.}
\label{fig_2}
\end{figure}
%===================================================  

Fig.~\ref{fig_1} illustrates the proposed mechanism of switching the polarity of RKKY in a F/S/F trilayer, with the outer ferromagnetic layers coupled ferro- or anti-ferromagnetically depending on temperature. In the conventional case of a uniform spacer (S) that is either nonmagnetic (N) or paramagnetic (P; for a strongly diluted magnetic component and having a low Curie point, $T_\text{C}$), the two outer F-layers experience AFM-RKKY at a suitable thickness of the spacer (Fig.~\ref{fig_1}(a)). On decreasing the temperature through $T_\text{C}$, the spacer becomes ferromagnetic (F*) and couples the outer F-layers parallel via \emph{direct exchange} (Fig.~\ref{fig_1}(b)). This mechanism allows to switch the magnetization state of the trilayer using the interplay of indirect (RKKY) and direct (non-RKKY) interactions in the structure. The disandvantage here is the strong proximity effect in the uniform spacer \cite{r27}, which significantly broadens the magnetic transition. 

Introducing a suitable composite spacer of type N/P(F*)/N as illustrated in Fig.~\ref{fig_1}(c,d) diminishes the direct exchange at the F/P interfaces and changes the mechanism for a thermo-magnetic transition in the system, based now fully on the sign reversal of the RKKY, due to a change in the effective thickness of the spacer driven by the Curie transition in F*. For $T < T_\text{C}$, the middle F*-layer of the N/F*/N spacer is ferromagnetic and couples to the outer F-layers by FM-RKKY (Fig.~\ref{fig_1}(d)), in contrast to the AFM-RKKY F-F coupling above $T_\text{C}$ (Fig.~\ref{fig_1}(c)). We demonstrate this new mechanism of RKKY polarity switching using Fe-Cr-based multilayers.

%------------------------------- II - Samples and Experimental details -----------------------------
%===================================================================================================
The multilayers were deposited at room temperature onto Ar pre-etched un-doped Si~(100) substrates using a dc-magnetron sputtering system (by AJA International). Layers of diluted Fe$_x$Cr$_{100-x}$ binary alloys of varied composition were deposited using co-sputtering from separate Fe and Cr targets. The alloy composition was controlled by setting the deposition rates of the individual Fe and Cr components, with suitable calibrations obtained by subsequent thickness profilometry. The magnetic properties were characterized using longitudinal magneto-optical Kerr effect (MOKE) measurements, carried our in the temperature range of 77-460~K using a MOKE setup equipped with an optical cryostat (by Oxford Instruments). Additionally, room-temperature magnetic characterization was performed using a vibrating-sample magnetometer (VSM, by Lakeshore Cryogenics).

%------------------------------ III - Experimental Results -----------------------------------------
%===================================================================================================

The classical, nominally nonmagnetic RKKY spacer of pure Cr was diluted with ferromagnetic Fe of concentration $x~=~0–30$~at.~\%~Fe and, in a subset of samples, subsequently structured into a tri-layered spacer with thermally variable ferromagnetic/paramagnetic characteristics. The Fe$_x$Cr$_{100-x}$ binary alloy is known to have good elemental solubility at low concentrations $x$ due to the similarity of lattice parameters of bulk Cr and Fe, with the Fe atoms being strongly magnetic even for low concentrations \cite{r22}. The critical temperature of the magnetic order-disorder transition of the bulk Fe-Cr alloy (its Curie temperature) is below room temperature for concentrations less than $x \approx 35$~at.~\%~Fe \cite{r23}. We note here and detail later that the ferromagnetic proximity effect in thin-film multilayers can significantly affect the actual critical temperature of the alloy.

The room-temperature magnetization-vs-field (M-B) data for Fe/Fe-Cr/Fe trilayers with uniformly diluted spacers, Fe$_x$Cr$_{100-x}$(1.5~nm), are shown in Fig.~\ref{fig_2}(a) (here and throughout the text the numbers in parenthesis correspond to the layer thickness in nm, unless noted otherwise). The M-B loop for Fe/Cr(1.5)/Fe ($x = 0~\%$) exhibits a strong indirect AFM coupling, indicated by the zero remnant magnetization and the high saturation field, $B_\text{s}\approx 100$~mT. Doping the Cr spacer with 5~\% or more of Fe gradually decreases $B_\text{s}$, eventually resulting in a single rectangular loop for $x \geqslant 15~\%$. Reference samples of Fe/Fe-Cr/Fe/Ir$_{20}$Mn$_{80}$ (not shown) with one of the outer Fe layers exchange-pinned by antiferromagnetic Ir$_{20}$Mn$_{80}$, reveal a single loop for $x \geqslant 15~\%$, shifted in field toward the pinning direction, indicating that for this concentration the outer Fe layers are coupled by direct exchange through the uniform Fe-Cr spacer.

%##################### Figure 3 #####################
%===================================================
\begin{figure}%[c]
\includegraphics[width=0.9\linewidth]{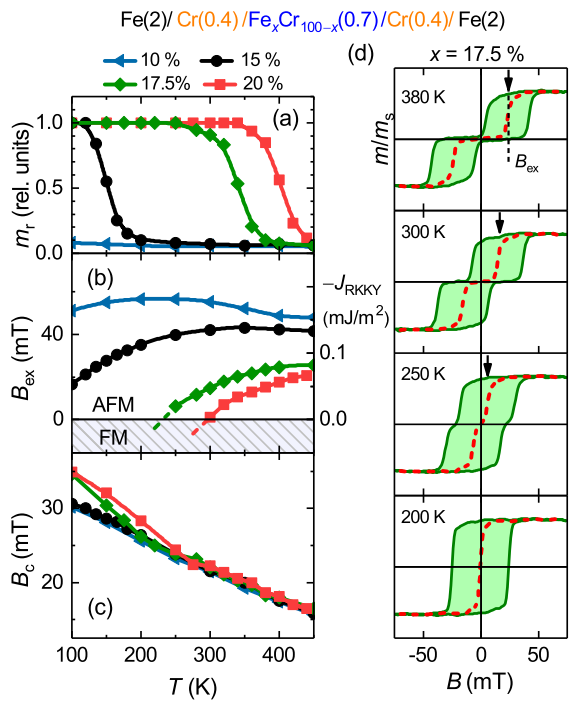}
\caption{Temperature dependence of (a) normalized remanent magnetic moment ($m_\text{r}$), (b) exchange field ($B_\text{ex}$) and the corresponding RKKY coupling strength $J_\text{RKKY}$ (right scale), and (c) minor-loop coercivity ($B_\text{c}$) for F/S/F trilayers with S~= Cr(0.4)/Fe$_x$Cr$_{100-x}$(0.7)/Cr(0.4), $x = 15, 17.5$ and 20~\%; (d) M-B loops for $x = 17.5~\%$ for representative temperatures above and below the effective Curie point of the gradient spacer (dashed lines indicate the center of the loops). Data obtained using MOKE magnetometry.}
\label{fig_3}
\end{figure}
%===================================================

In trilayers with composite RKKY/FM spacers, Cr($d_\text{Cr}$)/Fe$_x$Cr$_{100-x}$($d$)/Cr($d_\text{Cr}$), the direct exchange between the outer Fe layers and the inner-spacer layer Fe$_x$Cr$_{100-x}$ is suppressed by the thin nonmagnetic Cr layers at the respective Fe/Cr/Fe$_x$Cr$_{100-x}$ interfaces. Fig.~\ref{fig_2}(b) shows the relevant room-temperature M-B data for Fe/Cr(0.4)/Fe$_x$Cr$_{100-x}$(0.7)/Cr(0.4)/Fe. In contrast to the structures with uniformly diluted spacers (Fig.~\ref{fig_2}(a)), the trilayers with gradient spacers and low Fe-concentration ($x < 10~\%$) show essentially the same behavior to the classical RKKY tri-layer with pure-Cr spacers (Fe/Cr/Fe). As the Fe-concentration $x$ is increased above 10~\%, the M-B loops show a gradual decrease in the saturation field, still with pronounced RKKY at $x = 15~\%$, eventually resulting in a single rectangular loop at $x \geqslant 20~\%$ (uniformly diluted spacers show no RKKY already at $x = 15~\%$). Calibration experiments on exchange pinned Fe/Cr/Fe-Cr/Cr/Fe/IrMn samples (not shown) indeed revealed a field-offset single rectangular loop for $x \geqslant 20~\%$, confirming ferromagnetic coupling of the outer Fe layers via sequential RKKY/direct-exchange/RKKY interaction, with the nonmagnetic pure Cr layers breaking the direct exchange while mediating RKKY (FM-RKKY, expected for this small thickness of 0.4~nm, rather than AFM-RKKY).

Figs.~\ref{fig_3}(a)-(b) show the temperature evolution of the remanent magnetic moment, $m_\text{r}$, and the effective exchange field characterizing the strength of the RKKY interaction, $B_\text{ex}$, commonly taken to be the center point of the magnetic transition to saturation as defined graphically in Fig.~\ref{fig_3}(d) (top panel) with red dashed lines, centered at the respective magnetic transitions, excluding coercive effects intrinsic to the individual outer Fe layers. The RKKY exchange coupling strength was obtained in the standard way as $J_\text{RKKY}= M_\text{Fe}B_\text{ex}d_\text{Fe}$. A clear thermo-magnetic transition is seen in $m_\text{r}(T)$, from a fully saturated parallel state at low temperature to an antiparallel state with zero remanence at higher temperature, driven by the Curie transition in the inner-core of the spacer layer (F*). The same thermal transition is seen in the RKKY coupling strength: $B_\text{ex}$ ($J_\text{RKKY}$) is finite at higher temperature where the AFM-RKKY dominates the Fe-Fe IEC, and, for suitable ferromagnetic dilution of the spacer (17.5~\% and 20~\%), vanishes to zero at lower temperatures, indicating effectively ferromagnetic IEC. At the same time, the temperature dependences of the Fe layers' coercivity shown in Fig.~\ref{fig_3}(c) for all studied Fe-compositions and a fixed Cr-layer thickness of 0.4 nm is monotonous and featureless in the temperature range of the magnetic phase transition observed in $m_\text{r}$ and $B_\text{ex}$. $B_\text{c}$ is, as expected for single Fe films, increases somewhat at low temperatures, essentially insensitive to the specific composition and hence the magnetic state of the spacer. This indicates that the sharp thermo-magnetic transition observed originates from changes in the properties of the spacer layer.

%------------------------------ IV - Discussion ----------------------------------------------------
%===================================================================================================

The observed thermo-magnetic transition from ferro- to antiferromagnetic IEC with increasing temperature is consistent with the RKKY-switching mechanism designed into the structure and illustrated in Fig.~\ref{fig_1}. The dilute ferromagnetic aloy layer centered within the spacer is magnetically ordered at $T < T_\text{C}^*$ and, therefore, can effectively reflect spin-polarized electrons at the Cr/Fe$_x$Cr$_{100-x}$ interfaces, which in turn reflect from the Fe/Cr interfaces of the outer FM layers producing a spin density wave in Cr. For the small Cr layer thickness chosen (4--5~\AA), these spin-dependent conduction-electron reflections in Cr result in \emph{ferromagnetic} RKKY interactions across the individual pure-Cr layers. This FM-RKKY exchange interactions couple sequentially and \emph{ferromagnetically} via the thin center Fe-Cr layer in its ordered state. At $T_\text{C}^*$, the Fe-Cr layer becomes magnetically disordered, which diminishes the spin-dependent reflections responsible for the FM-RKKY in the structure, at the same time making the Fe-Cr layer to a large degree transparent to spin polarized current flowing between the outer-Fe/Cr interfaces. The total spacer thickness (15~\AA) is chosen such that the resulting RKKY at and above $T_\text{C}^*$ is antiferromagnetic. The increased spin depolarization due to now paramagnetic Fe in the thin Fe-Cr layer is insufficient to suppress the AFM-RKKY, which becomes progressively stronger as the temperature is increased.

%##################### Figure 4 #####################
%===================================================
\begin{figure}%[c]
\includegraphics[width=0.8\linewidth]{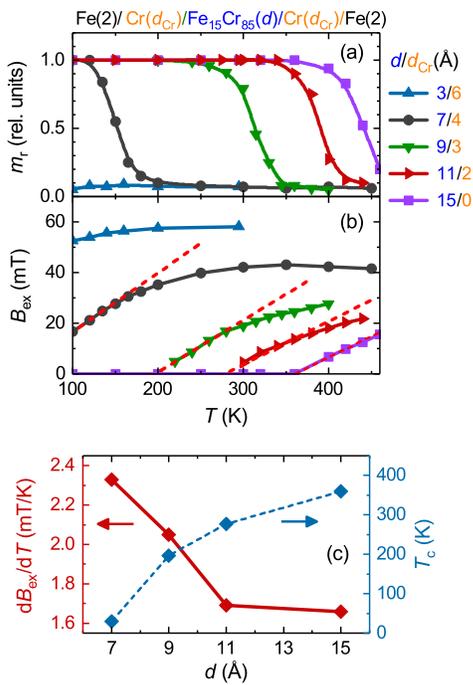}
\caption{Temperature dependence of (a) normalized remanent magnetization and (b) exchange field for F/S/F trilayers with S~= Cr($d_\text{Cr}$)/Fe$_{15}$Cr$_{85}$($d$)/Cr($d_\text{Cr}$), $d(d_\text{Cr}) =$ 3(6), 7(4), 9(3), 11(2) and 15(0)~\AA. (c) Initial slope of the $B_\text{ex}(T)$ dependence and critical temperature, $T_\text{C}^*$, vs. thickness $d$ for the spacer with constant concentration $x = 15~\%$.}
\label{fig_4}
\end{figure}
%===================================================

To further elucidate the mechanism of the RKKY-switching in our system, Fig.~\ref{fig_4} presents a set of data for structures with the spacer of type Cr($d_\text{Cr}$)/Fe$_{15}$Cr$_{85}$($d$)/Cr($d_\text{Cr}$), where the Fe-concentration of the inner-layer and the total thickness of the spacer were kept constant ($x = 15~\%, d_\text{tot} = 15$~\AA). These data show that, in addition to the strong dependence of $T_\text{C}^*$ on $x$ (Fig.~\ref{fig_3}), there is a pronounced dependence of the spacer's Curie point on the thickness of the Cr-layers, $d_\text{Cr}$ for a given $x =$~const. $d(d_\text{Cr}) = 15(0)$~\AA ~corresponds to the uniform-spacer case, with direct-exchange IEC (non-RKKY) at $T < T_\text{C}^*$ and the highest $T_\text{C}^*$ of the series. Introducing a thin Cr layer at the Fe/Fe$_{15}$Cr$_{85}$ interface lowers the $T_\text{C}^*$ and, at the same time, sharpens the thermo-magnetic transition in the structure, quantified in Fig.~\ref{fig_4}(b,c) as the initial slope of $B_\text{ex}(T)$. This slope of $B_\text{ex}(T)$ vs. $d$ shows a sharp increase for $d < 11 $ ~\AA, indicating a significantly more uniform magnetization profile in the inner-spacer due to the suppression of the direct-exchange proximity effect at the Fe/Fe-Cr interfaces. This variation is expected as our pure-Cr layers of 3~\AA~ and thicker are essentially continuous and transmit vanishing direct exchange, witnessed by the $T_\text{C}$ values approaching those in the bulk (near zero for relevant concentrations). We thus can conclude that in the structures with $d = 11 \text{ and } 15$~\AA~the Cr layers are too thin (0-2~\AA), potentially not continuous, which results in the relatively strong direct exchange through the spacer as well as the accompanying magnetic proximity effect, which broadens the FM-AFM transition and increases the effective $T_\text{C}^*$ (as discussed in, e.g., Refs.~\cite{r26,r27}). In contrast, the structures with $d < 9$~\AA~ ($d_\text{Cr} > 3$~\AA) show significantly to fully suppressed direct exchange between Fe and Fe-Cr below $T_\text{C}^*$, so the thermo-magnetic transition in the system is governed by competing FM-RKKY/direct-exchange/FM-RKKY and AFM-RKKY through the gradient spacer, with the balance shifting back and forth across the spacer's Curie point. 

It should be noted that magnetic ordering in the diluted ferromagnetic spacer used in this work may be affected by the \emph{indirect} RKKY proximity effect \cite{r28,r29}, nominally a function of $d_\text{Cr}$, which, however, should be significantly weaker that the direct-exchange proximity effect our structures are designed to avoid. An important related comment, firmly based on our data discussed above, is that the vanishing RKKY on uniform magnetic dilution of spacers reported in \cite{r19,r20} was likely due to a strong exchange-proximity effect and the accompanying direct-exchange coupling across the spacer, rather than potential Fermi-surface effects used as the interpretation.

%------------------------------ V - Conclusions -----------------------------
%===================================================================================================

In summary, we demonstrate thermal switching of the sign of indirect exchange in magnetic multilayers, where the spacer is a heterostructure with non-uniform magnetic dilution. Two designs with principally different magnetic ordering mechanisms are considered and contrasted. For the design with a uniform spacer, thermal switching is from a state of ferromagnetic IEC due to \emph{direct exchange} into a state of antiferromagnetic RKKY exchange as the temperature is increased over the effective $T_\text{C}^*$ of the alloyed spacer. For the composite-spacer design, where the direct interlayer exchange channel is suppressed by the pure-Cr spacers, heating induces a FM-RKKY to AFM-RKKY transition, driven by the Curie transition in the spacer's core, designed to result in a specific change of the effective thickness of the spacer. The relatively strong IEC ($\sim0.1$ mJ/m$^2$) and sharp FM-to-AFM RKKY transitions ($\sim10$ K) appear quite sufficient functionally to be interesting for applications in thermally-assisted spintronic devices, such as memory \cite{MRAM-Crocus-Prejbeanu_2013,Slonczewski_2010} and oscillators \cite{Kadigrobov_2010}.

%------------------------------ Acknowledgements -----------------------------
%===================================================================================================
 
Support from the Swedish Research Council (VR Grant No. 2014-4548) and the Swedish Stiftelse Olle Engkvist Byggm\"astare is gratefully acknowledged.

\bibliography{References}% Produces the bibliography via BibTeX.

\end{document}